\documentclass[sigconf]{acmart}
\pdfoutput=1
	
\newcommand{\model}{\mbox{$\mathop{\mathtt{SlimLogR}}\limits$}\xspace}
\newcommand{\name}{joint sparse linear recommendation and logistic regression model\xspace}
\newcommand{\SLIM}{\mbox{$\mathop{\mathtt{SLIM}}\limits$}\xspace}
\newcommand{\LogR}{\mbox{$\mathop{\mathtt{LogR}}\limits$}\xspace}
\newcommand{\modelIn}{\mbox{$\mathop{\model_{\mathtt{in}}}\limits$}\xspace}
\newcommand{\modelEx}{\mbox{$\mathop{\model_{\mathtt{ex}}}\limits$}\xspace}
\newcommand{\Rand}{\mbox{$\mathop{\mathtt{Rand}}\limits$}\xspace}
\newcommand{\LROnly}{\mbox{$\mathop{\mathtt{LogR}}\limits$}\xspace}
\newcommand{\SLIMOnly}{\mbox{$\mathop{\mathtt{SLIM}}\limits$}\xspace}
\newcommand{\Sep}{\mbox{$\mathop{\SLIM\text{+}\LogR}\limits$}\xspace}
\newcommand{\predBin}{\mbox{$\mathop{\LogR_{\mathtt{b}}}\limits$}\xspace}
\newcommand{\predScr}{\mbox{$\mathop{\LogR_{\mathtt{s}}}\limits$}\xspace}

\newcommand{\rec}{\mbox{$\mathop{\mathtt{rec_{t}}}\limits$}\xspace}
\newcommand{\recall}{\mbox{$\mathop{\overline{\mathtt{rec_{t}}}}\limits$}\xspace}
\newcommand{\precn}{\mbox{$\mathop{\mathtt{prec_{t}}}\limits$}\xspace}
\newcommand{\precision}{\mbox{$\mathop{\overline{\mathtt{{prec}_t}}}\limits$}\xspace}
\newcommand{\acc}{\mbox{$\mathop{\mathtt{acc_{t}}}\limits$}\xspace}
\newcommand{\accuracy}{\mbox{$\mathop{\overline{\mathtt{{acc}_t}}}\limits$}\xspace}

\newcommand{\drug}{\mbox{$\mathop{\mathnormal{d}}\limits$}\xspace}
\newcommand{\D}{\mbox{$\mathop{\boldsymbol{a}}\limits$}\xspace}

\newcommand{\DataPoll}{\mbox{$\mathop{A_{pool}}\limits$}\xspace}

\newcommand{\OR}{\mbox{$\mathop{\mathtt{OR}}\limits$}\xspace}
\newcommand{\ORs}{\mbox{$\mathop{\mathtt{OR}}\limits$s}\xspace}

\newcommand{\adr}{\mbox{$\mathop{\text{ADR}}\limits$}\xspace}

\newcommand{\Dall}{\mbox{$\mathop{A_{\text{FAERS}}}\limits$}\xspace} 
\newcommand{\Dataset}{\mbox{$\mathop{A_{*}}\limits$}\xspace}
\newcommand{\Dtrain}{\mbox{$\mathop{A_{*trn}}\limits$}\xspace}
\newcommand{\Dtest}{\mbox{$\mathop{A_{*tst}}\limits$}\xspace}

\newcommand{\Nneg}{\mbox{$\mathop{A^-}\limits$}\xspace}  
\newcommand{\Nminus}{\mbox{$\mathop{A_-^-}\limits$}\xspace}  
\newcommand{\Nzero}{\mbox{$\mathop{A_0^-}\limits$}\xspace}  
\newcommand{\Mpos}{\mbox{$\mathop{A^+}\limits$}\xspace}  
\newcommand{\Mzero}{\mbox{$\mathop{A^+_0}\limits$}\xspace}  
\newcommand{\Mplus}{\mbox{$\mathop{A^+_+}\limits$}\xspace}  

\usepackage{amsthm}
\usepackage{comment,enumitem,booktabs}
\usepackage{graphicx}
\usepackage{wrapfig}
\usepackage{amsmath}
\usepackage{amsfonts}
\usepackage{amssymb}
\usepackage{mathtools}
\usepackage{array}
\usepackage{algorithm}
\usepackage{mathtools,nccmath}

\usepackage{algpseudocode,algorithmicx}

\usepackage{algpseudocode}
\usepackage{pgfplotstable}

\usepackage{booktabs}
\setlist{nolistsep}

\usepackage[misc]{ifsym}
\usepackage{centernot}
\usepackage{longtable}
\usepackage{wrapfig}
\usepackage{xr}
\usepackage{tikz}
\usepackage{./style/optidef}
\usepackage{./style/slashbox}
\usetikzlibrary[automata, positioning]
\usepackage[flushleft]{threeparttable}
\usepackage{multirow}
\usepackage{caption}
\usepackage{subcaption}
\usepackage{xspace}
\usepackage[figuresright]{rotating}
\usepackage{cellspace}
\usepackage[table]{xcolor}

\graphicspath{
{./figure/}
{./figure/convergence/}
}

\begin{document}

\setlength{\abovedisplayskip}{3pt}
\setlength{\belowdisplayskip}{3pt}

\settopmatter{printacmref=false} 
\title{Drug Recommendation toward Safe Polypharmacy}


\author{Wen-Hao Chiang}
\affiliation{%
  \institution{Department of Computer \& Information Science\\
  Indiana University - Purdue University Indianapolis}
  \city{Indianapolis}
  \state{Indiana}
  \postcode{46202}
}
\email{chiangwe@iupui.edu}

\author{Li Shen}
\affiliation{%
  \institution{Department of Biostatistics, Epidemiology and Informatics\\
    University of Pennsylvania}
  \city{Philadelphia}
  \state{Pennsylvania}
  \postcode{19104}
}
\email{Li.Shen@pennmedicine.upenn.edu}

\author{Lang Li}
\affiliation{%
  \institution{Department of Biomedical Inforamtics\\
  Ohio State University}
  \city{Columbus}
  \state{Ohio}
  \postcode{43210}
}
\email{Lang.Li@osumc.edu}

\author{Xia Ning}
\authornote{Corresponding author.}
\affiliation{%
  \institution{Department of Computer \& Information Science\\
  Indiana University - Purdue University Indianapolis}
  \city{Indianapolis}
  \state{Indiana}
  \postcode{46202}
}
\email{xning@iupui.edu}

\begin{abstract}

Adverse drug reactions (ADRs) induced from high-order drug-drug interactions (DDIs)
due to polypharmacy represent a significant public health problem.
%
%
In this paper, we formally formulate the to-avoid
and safe (with respect to ADRs) drug recommendation problems when
multiple drugs have been taken simultaneously.
We develop a joint model with a recommendation component and an ADR label
prediction component to recommend for a prescription a set of to-avoid drugs that
will induce ADRs if taken together with the prescription.
We also develop real drug-drug interaction datasets and corresponding evaluation
protocols. 
Our experimental results on real datasets demonstrate the strong performance
of the joint model compared to other baseline methods. 

\end{abstract}

\ccsdesc[500]{Computing methodologies~Machine learning}
\ccsdesc[500]{Computing methodologies~Supervised learning}
\ccsdesc[500]{Applied computing~Health care information systems}
\ccsdesc[500]{Applied computing~Health informatics}
\ccsdesc[500]{Applied computing~Bioinformatics}
\ccsdesc[300]{Information systems~Data mining}

\keywords{Adverse Drug Reaction, Polypharmacy, Drug Recommendation}

\maketitle


\section{Introduction}
\label{sec:intro}



Adverse drug reactions (ADRs) induced from high-order drug-drug interactions (DDIs)
due to polypharmacy (i.e., simultaneous use of multiple drugs) represent
a significant public health problem. 
ADRs refer to undesired or harmful reactions due to drug administration.
One of the major causes for ADRs is DDIs, which happen when the pharmacological
effects of a drug are altered by the actions of another drug,
leading to unpredictable clinical consequences.
DDIs were responsible for approximately 26\% of the ADRs,
affected 50\% of inpatients~\cite{Ramirez2010},
and caused nearly 74,000 emergency room visits and
195,000 hospitalizations annually in the US~\cite{Percha2013}.
The increasing popularity of polypharmacy continues to expose
a significant and growing portion of the population to unknown or poorly understood DDIs
and associated ADRs.
The National Health and Nutrition Examination Survey~\cite{nhn} reports that more
than 76\% of the elderly Americans take two or more drugs every day.
Another study~\cite{Iyer2014} estimates that
about 29.4\% of elderly American patients take six or more drugs every day.
This has made understanding high-order DDIs and their associated
ADRs urgent and important. 

Current research on DDIs and their associated ADRs is primarily focused on mining
and detecting DDIs for knowledge discovery (literature review in
Section~\ref{sec:review:ddi}). 
In applying the knowledge in practice for preventive, proactive and
person-centered healthcare, it also requires predictive power to deal with unknowns
and uncertainties, and to provide evidence-based suggestions to facilitate future drug
use.
As wet-lab based experimental validation scheme for DDI study
still falls behind due to its low throughput and lack of scalability, but meanwhile
huge amounts of electronic medical record data become increasingly available, 
data-driven computational methodologies appear more appealing to tackle DDI and ADR
problems. 
Unfortunately, computational effects to facilitate future polypharmacy, particularly to
assist safe multi-drug prescription, are still in their infancy.

In this paper, we present a computational method toward the goal of assisting future safe
multi-drug prescriptions.
Here we use the term ``prescription'' to represent a set of drugs that have been
taken together, even though there could be non-prescription drugs.
Thus, we tackle the following \emph{to-avoid drug recommendation} problem
and \emph{safe drug recommendation} problem. 
\newtheorem{mydef}{Definition}
\begin{mydef}
  \underline{To-avoid Drug Recommendation}:
  given the multiple drugs
  in a prescription that have been taken simultaneously, recommend a short ranked list of
  drugs that should be avoided taking together with the prescription in order to avoid a
  particular ADR. 
\end{mydef}
\begin{mydef}
  \underline{Safe Drug Recommendation}:
  given the multiple drugs
  in a prescription that have been taken simultaneously, recommend a short ranked list of
  safe drugs that, if taken together with the prescription, are not likely to induce a
  particular ADR. 
\end{mydef}
Please note that in this study we only consider one ADR and thus drug safety is
only considered with respect to that ADR. 
To the best of our knowledge, this is the first work to formally formulate the above
problems and provide a computational solution framework.

The above two recommendation problems are significant particularly for healthcare
practice.
They can enable evidence-based suggestions to help polypharmacy
decision making, and induce novel hypotheses on new high-order DDIs and associated
ADRs.
These problems are different from traditional product recommendation problems in
e-commerce, in which no label information is involved on a set of
user's previous purchased products as a whole (each product is analogous to a drug;
a set of previous purchased products is analogous to a prescription).
In addition, ADRs induced from a set of drugs could be due to pharmaceutical,
pharmacokinetic or pharmacodynamic interactions among only a (unknown) subset of the
drugs. In e-commerce recommendation problems, the notion of subset synergy and
interactions is very weak.
Due to the above reasons, the to-avoid drug recommendation and safe drug recommendation
problems are inherently very non-trivial.

Our contributions to solving the two new drug recommendation problems are summarized as
follows:
\begin{itemize}[noitemsep, leftmargin=*]
\item We developed a \name (\model),
  with a drug recommendation component and an ADR label
  prediction component (Section~\ref{sec:method}).
  The recommendation component captures drug co-prescription
  patterns among ADR and non-ADR inducing prescriptions,
  respectively, and uses such patterns to recommend drugs.
  The ADR label prediction component re-ranks all the recommended drugs based on
  their ADR probabilities to produce final recommendations. 
  These two components are learned concurrently in \model so that the recommended
  drugs are more likely to introduce expected ADR labels.
\item We developed a protocol to mine high-order DDIs and their associated ADRs,
  and provided a DDI dataset to the public~\footnote{The dataset will be publicly available
    upon the acceptance of this paper} (Section~\ref{sec:materials}).
\item We developed new evaluation protocols and evaluation metrics to evaluate the
  performance of drug recommendation (Section~\ref{sec:evaluation}).
\item We conducted comprehensive experiments to evaluate \model, and provided
  case study on recommended drugs (Section~\ref{sec:exp}). Our experiments demonstrate
  strong performance of \model compared to other methods. 
\end{itemize}

\section{Related Work}
\label{sec:review}

\subsection{DDI and ADR Study}
\label{sec:review:ddi}

Significant research efforts have focused on detecting DDIs. 
These efforts can be broadly classified into four categories~\cite{Vilar2014}. 
Methods of the first category analyze medical literature and/or electronic medical records,
and extract mentioned drug pairs~\cite{Iyer2014}. 
Methods in the second category integrate various biochemical and molecular
drug/target data to measure drug-drug similarities
and score/predict pairwise DDIs.
%
These data include target information~\cite{Vilar2014},
phenotypic and genomic information~\cite{Cheng2014}, and drug side effects~\cite{Tatonetti2011}, etc.
Methods of the third category leverage healthcare information on social media and
  online communities to detect DDIs~\cite{Yang2013}. 
  The fourth category focuses on using numerical models to predict the dose responses
  to multiple drugs~\cite{Chiang2018}. 
A recent thread is dedicated to understanding the interaction patterns
among high-order DDIs, and how such patterns can relate to induced ADRs~\cite{Ning2017}.
%
%
%
%
%

\subsection{Recommender Systems}
\label{sec:review:rs}

Top-$N$ recommender systems, which recommend the top-$N$ items that are most likely to be
preferred by users, have been used in a variety of applications in e-commerce, social
networking and healthcare, etc.
The methods can be broadly classified in two categories. 
The first category is the neighborhood-based collaborative filtering methods~\cite{Ning2015},
which produce recommendations based on similar users or similar items. 
The second category is model-based methods, particularly
latent factor models~\cite{Ricci2011} that 
%
factorize the user-item matrix
into (low-rank) user factors and item factors,
and predict user preference over items using the latent factors. 
%
%
Recent recommender systems have also been advanced by the significant contribution
from deep learning~\cite{Zhang2017}, where user preferences and item characteristics
can be learned in deep architectures.
On the other hand, recommender systems methods have also been integrated with
regression methods to tackle joint problems of missing value recovery and
prediction~\cite{Lan2016}. 

Recommender systems have been recently applied to prioritizing healthcare information,
due to a tremendous need for personalized healthcare~\cite{Pfeifer2014}.
Current applications along this line include recommending physicians to patients on
specific diseases~\cite{Guo2016}, 
recommending drugs for patient symptoms~\cite{Wu2015}, 
recommending nursing care plans~\cite{Duan2011}, and
therapy decision recommendation~\cite{Grasser2017}.
However, to the best of our knowledge, very little research has been done on
new prescription recommendation particularly with ADR concerns.

\section{Definitions and Notations}
\label{sec:def}

%
In this paper, we use \drug to represent a drug, 
and a binary matrix $A  \in \mathbb{R}^{m \times n}$ to represent prescription data.
Each row of $A$ represents a prescription, denoted as $\boldsymbol{a}_i$ ($i = 1, \cdots, m$),
and each column of $A$ corresponds to a drug. Thus,
there are $m$ prescriptions and $n$ drugs in total represented by $A$. 
If a prescription $\boldsymbol{a}_i$ contains a drug $\drug_j$, the $j$-th entry in $\boldsymbol{a}_i$
(i.e., $a_{i,j}$) will be 1, otherwise 0.
Thus, $\boldsymbol{a}_i$
is a binary row vector representing the existence of drugs in the $i$-th prescription.
%
%
Note that all the prescriptions are unique in $A$. Drug dosage in prescriptions
is not considered in this study.
When no ambiguity arises, the terms ``a prescription'' and ``a binary vector $\boldsymbol{a}$''
are used exchangeably,
and ``a set of prescriptions'' and ``a matrix $A$'' are also used exchangeably.
In addition, 
a label $y_i$ is assigned to $\boldsymbol{a}_{i}$ to indicate whether
$\boldsymbol{a}_i$ induces a certain ADR (denoted as $y_i=1$ or $y_i^+$)
or not (denoted as $y_i=-1$ or $y_i^-$).
In this paper, we only consider one ADR and thus the labels are binary. 
In addition, superscript $^+$ indicates information/data related to ADR
induction (i.e., positive information), superscript $^-$ indicates information/data related
to no ADR induction (i.e., negative information) and accent  $\tilde{A}$ indicates
estimated $A$. 
%
%

%

\section{Background}
\label{sec:background}

\subsection{\SLIM for top-$N$ Recommendation}
\label{sec:background:SLIM}

%
Sparse Linear Method (\SLIM)~\cite{Ning2011} is an efficient and state-of-the-art algorithm for
top-$N$ recommendation that was initially designed for e-commerce applications.
In the drug recommendation problem, given a drug prescription $\boldsymbol{a}_i$,
\SLIM models the score of how likely an additional drug $\drug_j$ should be
co-prescribed with $\boldsymbol{a}_i$
as a sparse linear aggregation of the drugs in $\boldsymbol{a}_i$, that is, 
\begin{equation}
  \label{eqn:RecScore}
    \tilde{a}_{ij} = \boldsymbol{a}_i\boldsymbol{w}_j^{\intercal},
\end{equation}
where $\tilde{a}_{ij}$ is the estimated score of $\drug_j$ in $\boldsymbol{a}_i$ (can have values rather
than 0 and 1),
and $\boldsymbol{w}_{j}^{\intercal}$ is a sparse column vector of aggregation coefficients.
Note that $a_{ij}=0$, that is, $\drug_j$ is not included $\boldsymbol{a}_i$ originally. 
Drugs with high scores calculated as above will be recommended to the prescription.
Thus, the scores are referred to as recommendation scores, and a prescription composed of
$\boldsymbol{a}_i$ and a recommended drug $\drug_j$ is referred to as a new prescription with respect to
$\boldsymbol{a}_i$, denoted as $\boldsymbol{a}_i \cup \{\drug_j\}$.
The intuition of using \SLIM for to-avoid drug recommendation will be discussed later
in Section~\ref{sec:method:model:slim}. 

%
%

To learn $W = [\boldsymbol{w}_1^{\intercal}, \boldsymbol{w}_2^{\intercal}, \cdots, \boldsymbol{w}_n^{\intercal}]$,
\SLIM solves the following optimization problem, 
\begin{eqnarray}
  \label{opt:SLIM}
  \begin{aligned}
    & \underset{W}{\min} && \hspace{-20pt}\SLIM(A; W, \alpha, \lambda)  = \frac{1}{2} \|{A - AW}\|^2_F +
    \frac{\alpha}{2} \|{W}\|^2_F + \lambda \|{W}\|_{\ell_1}, \\
    & \text{subject to}    	 && {W}\geq 0, \text{diag}(W) = 0,  \\
  \end{aligned}
\end{eqnarray}
where $\|{W}\|_{\ell_1} = \sum_{i=1}^{n} \sum_{i=1}^{n} |w_{ij}|$ is the entry-wise $\ell_{1}$-norm 
of $W$, and $\|{\cdot}\|_{F}$ is the matrix Frobenius norm.
The \SLIM formulation can also be considered as a way to reconstruct $A$ by itself through learning
the patterns (i.e., $W$) in $A$. 
Therefore, $W$ converts a binary $A$ into its estimation $\tilde{A}$ of floating values,
which cover
recover unseen non-zero entries in $A$.

%
%

\subsection{Logistic Regression for Label Prediction}
\label{sec:background:LogR}

We can formulate the problem of predicting whether a prescription of multiple drugs induces a particular 
ADR as a binary classification problem, and solve the problem using
logistic regression (\LogR). 
In \LogR, the probability of a prescription $\boldsymbol{a}_i$ inducing the ADR
is modeled as follows, 
\begin{equation}
  \label{eqn:Probability}
      p( y_{i} | \boldsymbol{a}_{i}; \boldsymbol{x}, c) =  (1 + \text{exp}(-y_{i} (\boldsymbol{a}_{i}\boldsymbol{x}^{\intercal}+c)))^{-1},
\end{equation}
where $\boldsymbol{x}^{\intercal}$ and $c$ are the parameters.
To learn the parameters, \LogR solves the following optimization problem,
\begin{eqnarray}
  \label{opt:LogR}
  \begin{aligned}
    & \underset{{\boldsymbol{x},c}}{\min} && \LogR(\boldsymbol{y}|A; \boldsymbol{x}, c, \beta, \gamma)\\
    &                                     && = {\sum}_{i=1}^{m} \text{log}\{1+\text{exp}[-y_{i}(\boldsymbol{a}_{i}\boldsymbol{x}^{\intercal}+c)]\} +
    \frac{\beta}{2}\|{\boldsymbol{x}}\|^2_2 + \gamma\|{\boldsymbol{x}}\|_{1}, \vspace{-10pt}\\    
  \end{aligned}
\end{eqnarray}
where $\boldsymbol{y} = [y_1; y_2; \cdots, y_m]$,
$\|{\boldsymbol{x}}\|_{1} = \sum_{i=1}^{n}|x_{i}|$, 
and $\|{\boldsymbol{x}}\|_{2}^2 = \sum_{i=1}^{n}{{x_{i}}^2}$. 
%

\section{Drug Recommendation}
\label{sec:method}

\subsection{Joint Sparse Linear Recommendation and Logistic Regression Model}
\label{sec:method:model}

%
%
We propose a novel model to conduct the task of to-avoid and safe drug
recommendation to existing prescriptions.
The model recommends a set of additional to-avoid drugs to an existing prescription
such that each of the recommended drugs should not be taken together with the prescription
in order to avoid ADRs. The model also recommends safe drugs that can be taken together
with the prescription. 
This novel model consists of a drug recommendation component and an ADR prediction
component. The recommendation component is instantiated from
\SLIM (Section~\ref{sec:background:SLIM}), and the ADR prediction
component uses logistic regression \LogR (Section~\ref{sec:background:LogR}).
This model is referred to as \name and denoted as \model.

\model learns the \SLIM and \LogR components jointly through solving the following optimization
problem.
\begin{eqnarray}
  \label{opt:SLIMLRWsep}
  \begin{aligned}
    & \underset{W^{+},W^{-},~\boldsymbol{x},~c}{\min}
    && 
    \SLIM(A^+; W+, \alpha, \lambda) + \SLIM(A^-; W^-, \alpha, \lambda) \\    
    &
    && \omega \{
    \LogR(\boldsymbol{y}^+|\tilde{A}^+\circ M^+; \boldsymbol{x}, c, \beta, \gamma) + \\
    &
    && \textcolor{white}{\omega\{}
    \LogR(\boldsymbol{y}^-|\tilde{A}^-\circ M^-; \boldsymbol{x}, c, \beta, \gamma)\}\\
    & \text{subject to}
    && \tilde{A}^{+} = A^{+} W^{+}, 
       \tilde{A}^{-} = A^{-} W^{-}, \\
    &
    && W^{+} \geq 0, W^{-}\geq 0, 
    %
       \text{diag}(W^{+}) = 0, \text{diag}(W^{-}) = 0. \\
  \end{aligned}
\end{eqnarray}
%
%
In \model, $A^+$/$A^-$ is a set of training prescriptions known
to induce ADRs/not to induce ADRs, respectively,
and $\tilde{ A}^{+}$/$\tilde{ A}^{-}$ is the respective estimation
from a \SLIM model (Equation~\ref{eqn:RecScore}),
that is, \mbox{$\tilde{A}^{+} = A^{+} W^{+}$} and \mbox{$\tilde{A}^{-} = A^{-} W^{-}$},
where $W^+$ and $W^-$ are the \SLIM model parameters.
In addition, $M$ is a mask matrix that selects values in
$\tilde{A}$ in predicting $\boldsymbol{y}$ (i.e., the ADR labels on $A$ prescriptions)
in the
\LogR component, and $\circ$ is the entry-wise Hadamard product.
%

\subsubsection{Learning Co-Prescription Patterns via \SLIM}
\label{sec:method:model:slim}

\model learns drug co-prescription patterns using \SLIM
(\mbox{$\tilde{ \boldsymbol{a}} = \boldsymbol{a} W$}, $W \geq 0, \text{diag}(W) = 0$,
or explicitly, $\tilde{a}_{i,j} = \sum_{k, k \neq j }a_{i,k} W_{k,j}$), that is, 
whether a drug should be included in a prescription
is modeled as a linear function of other drugs that are included in the prescription.
This modeling scheme is motivated by the existence of strong co-prescription 
drug pairs as we observe from real data (Section~\ref{sec:exp:patterns}). 
%
%
Meanwhile, linearality is a simplistic relation to model the co-prescription patterns, in which
larger coefficients represent stronger co-prescription relation.
Note that the zero-diagonal constraint on the coefficient matrix $W$ in \SLIM
(i.e., $\text{diag}(W^+) = 0$, $\text{diag}(W^-) = 0$ in Equation~\ref{opt:SLIMLRWsep})
excludes the possibility that a drug is co-prescribed
with itself. 
In addition, the non-negativity constraint on $W$ (i.e., $W^{+} \geq 0$, $W^{-}\geq 0$ in
Equation~\ref{opt:SLIMLRWsep}) ensures that the learned relation is on drug co-appearance. 
%
The regularization term $\|W\|_{\ell_1}$ induces sparsity in $W$ because not all the drugs tend
to be co-prescribed. 

%

%
In \model, the patterns in ADR inducing and non-ADR inducing
prescriptions are modeled using two \SLIM models
(i.e., $\tilde{A}^{+} = A^{+} W^{+}$
and $\tilde{A}^{-} = A^{-} W^{-}$) because it is expected
these two patterns are different, and thus the prospectively
learned $W^+$ and $W^-$ will present different patterns.
The pattern difference is indicated during the data pre-processing
(Section~\ref{sec:materials:mining} and~\ref{sec:materials:training}, Table~\ref{table:Statis_whole}).
The $W$ coefficients in the \SLIM component are learned by minimizing the difference between
$A$ and $\tilde{A}$ estimated from \SLIM. Note that ${A}$ is a binary matrix 
but $\tilde{A}$ can have floating values rather than 0 and 1. 
Once the linear aggregation parameter $W$ is learned, it can be used to
recommend additional drugs
that are likely to be co-prescribed with an existing prescription.

\subsubsection{Predicting ADR via \LogR}
\label{sec:method:model:logreg}

\model uses \LogR 
to predict whether a prescription will induce ADRs or not
(i.e., $\sum_{i=1}^{m}\log \{1+\text{exp}[ -y_{i}((\tilde{\boldsymbol{a}}_{i} \circ \boldsymbol{m}_{i} ) \boldsymbol{x}^\intercal + c ) \}$).
\LogR produces a probability of ADR induction from
a linear combination of drugs in a prescription (Equation~\ref{eqn:Probability}),
and these probabilities can be further used to rank such prescriptions. 
Note that the estimated prescriptions $\tilde{A}$ from \SLIM
rather than $A$ is used to train \LogR.
This connects \SLIM and \LogR to further enforce that the \SLIM component
learns co-prescription patterns that better correlate with their ADR induction labels (i.e.,
$\boldsymbol{y}$ in Equation~\ref{opt:SLIMLRWsep}).
Meanwhile, the use of $\tilde{A}$ also generalizes the ability of \LogR to
predict for new prescriptions, as $\tilde{A}$ will have new drugs compared to
$A$ (as discussed in Section~\ref{sec:background:SLIM}).

In \LogR, $M$ selects values in $\tilde{A}$ in predicting $\boldsymbol{y}$. 
We have the following two approaches to selecting $M^{+}$ and $M^{-}$.
\emph{Inclusive \model (\modelIn)}:
In this approach, \mbox{$M = \boldsymbol{1}$} (i.e., a matrix of all 1's), and therefore,
all the drugs in $\tilde{A}$ and their values will be used in the \LogR component. 
The use of all these drugs plays the role to enforce that the \SLIM component
produces large recommendation scores on prescribed drugs in $A$ and small scores on un-prescribed drugs. 
In this way, 
the prescribed drugs will dominate the label prediction in the \LogR component in order to produce
accurate label prediction. 
%
This approach is referred to as inclusive \model and denoted as \modelIn. 

\emph{Exclusive \model (\modelEx)}: 
In this approach, \mbox{$M = \mathbb{I}(A)$} where $\mathbb{I}$
is an indicator function ($\mathbb(I)(x) = 0$ if $x = 0$, 1 otherwise),
that is, only the prescribed drugs in $A$ and their values
are used to ensure that \LogR is able to make accurate prediction. 
This approach is referred to as exclusive \model and denoted as \modelEx.

\subsection{Training \model}
\label{sec:method:training} 

The optimization problem \ref{opt:SLIMLRWsep} can be solved through 
the alternating direction method of multipliers (ADMM)~\cite{Boyd2011}.
We introduce a new variable $Z$ and thus the following augmented Lagrangian
as the new objective to optimize: 
%
%
%
 \begin{eqnarray*}
  \label{opt:SLIMLRWsepLM}
  \begin{aligned}
    &\underset{\substack{W^{+}, W^{-}, Z^+, Z^-, ~\boldsymbol{x}, ~c}}{\min}
    && L({W^{+}},{W^{-}},{Z^{+}},{Z^{-}},~\boldsymbol{x},~c,~\boldsymbol{u}^{+},~\boldsymbol{u}^{-},\rho^{+},\rho^{-}) = \\
    &
    &&
    \SLIM(A^+; W+, \alpha, \lambda) + \SLIM(A^-; W^-, \alpha, \lambda) \\    
        &
    && \omega \{
    \LogR(\boldsymbol{y}^+|\tilde{B}^+\circ M^+; \boldsymbol{x}, c, \beta, \gamma) + \\
    &
    && \textcolor{white}{\omega\{}
    \LogR(\boldsymbol{y}^-|\tilde{B}^-\circ M^-; \boldsymbol{x}, c, \beta, \gamma)\}\\
    &
    &&
    ~{\boldsymbol{u}^{+}}^{\intercal}\boldsymbol{v}^+ + 
    \frac{\rho^{+}}{2} \| \boldsymbol{v}^+ \|^2_2 +
    ~{\boldsymbol{u}^{-}}^{\intercal}\boldsymbol{v}^- + 
    \frac{\rho^{-}}{2} \| \boldsymbol{v}^- \|^2_2, \\	
    & \text{subject to}
    && \hspace{-25pt} \tilde{B}^{+} = A^{+} Z^{+}, 
    \tilde{B}^{-} = A^{-} Z^{-}, \\
    &
    && \hspace{-25pt} \boldsymbol{v}^+ = \text{vec}({W^{+}}) - \text{vec}({Z^{+}}),
    \boldsymbol{v}^- = \text{vec}({W^{-}}) - \text{vec}({Z^{-}}),\\
    &
    && \hspace{-25pt} W^+ = Z^+, W^- = Z^-, W^{+} \geq 0, W^{-}\geq 0, \\
    &
    && \hspace{-25pt} \text{diag}(W^{+}) = 0, \text{diag}(W^{-}) = 0, \\
  \end{aligned}
 \end{eqnarray*}
%
where ${\boldsymbol{u}^{+}}$ and ${\boldsymbol{u}^{-}}$ are the Lagrange multipliers;
$\rho^{+}, \rho^{-}$ > 0 are the penalty parameters,
and $\text{vec}(\cdot)$ is the vectorization of a matrix.
The algorithm to solve the problem is presented in Algorithm~\ref{alg:ADMMupdate}.
\begin{algorithm}
 \caption{Learning $\model$}
 \label{alg:ADMMupdate}
 \begin{algorithmic}[1]  
   \begin{small}
   \Function{\model}{$A$, $\omega$, $\alpha$, $\lambda$, $\beta$, $\gamma$}
   \State ${\rho^{+}} = 10,{\rho^{-}} = 10$, $\boldsymbol{u}^{+}_{(0)} = \boldsymbol{0}$, $\boldsymbol{u}^{-}_{(0)} = \boldsymbol{0}$, $k = 0$
   \State $Z^{+}_{(0)} = W^{+}_{(0)}$, $Z^{-}_{(0)} = W^{-}_{(0)}$
   \State learn $W^{+}_{(0)}$ and $W^{-}_{(0)}$ from \SLIMOnly (Section~\ref{sec:comparison:SLIMOnly}) 
   \State learn $\boldsymbol{x}_{(0)}$ and $c_{(0)}$ from \LROnly (Section~\ref{sec:Comparison:LROnly})
   \While{not converge}
   \State 
   \vspace*{-\baselineskip}
   \begin{fleqn}[\dimexpr\leftmargini-\labelsep]
     \setlength\belowdisplayskip{-5pt}
     \begin{equation*}
       \begin{multlined}[c]
         \quad\quad \{W^{+}_{(k+1)}, W^{-}_{(k+1)}\}\coloneqq
         \underset{ {W^{+}, W^{-}}}{\mathrm{argmin}} ~~ L( W^{+}_{(k)}, W^{-}_{(k)}, Z^{+}_{(k)}, Z^{-}_{(k)}, \vspace{-5pt}\\
         \textcolor{white}{\quad W^{+}_{(k+1)} \coloneqq}
         \textcolor{white}{\underset{ W^{+} }{\mathrm{argmin}} ~~}
         \quad\quad\quad\quad\boldsymbol{x}_{(k)},~c_{(k)},~\boldsymbol{u}^{+}_{(k)},~\boldsymbol{u}^{-}_{(k)})\\
       \end{multlined}
     \end{equation*}
     \vspace{-15pt}
   \end{fleqn}%
   \State
   \vspace*{-\baselineskip}
   \begin{fleqn}[\dimexpr\leftmargini-\labelsep]
     \setlength\belowdisplayskip{-5pt}
     \begin{equation*}
       \begin{multlined}[c]
         \quad\quad \{Z^{+}_{(k+1)}, Z^{-}_{(k+1)}\} \coloneqq
         \underset{ Z^{+}, Z^{-} }{\mathrm{argmin}} ~~ L( W^{+}_{(k+1)}, W^{-}_{(k+1)}, Z^{+}_{(k)}, Z^{-}_{(k)},\vspace{-5pt}\\
         \textcolor{white}{~\quad\quad Z^{+}_{(k+1)} \coloneqq}
         \textcolor{white}{\underset{ Z^{+} }{\mathrm{argmin}} ~~ }
         \quad\boldsymbol{x}_{(k)},~c_{(k)},~\boldsymbol{u}^{+}_{(k)},~\boldsymbol{u}^{-}_{(k)})\\
       \end{multlined}
     \end{equation*}
     \vspace{-15pt}
   \end{fleqn}%
   \State
   \vspace*{-\baselineskip}
   \begin{fleqn}[\dimexpr\leftmargini-\labelsep]
     \setlength\belowdisplayskip{-5pt}
     \begin{equation*}
       \begin{multlined}[c]
         \quad\quad\{~\boldsymbol{x}_{(k+1)},~c_{(k+1)}\} \coloneqq 
         \underset{\boldsymbol{x},~c}{\mathrm{argmin}}~~L( W^{+}_{(k+1)}, W^{-}_{(k+1)}, Z^{+}_{(k+1)}, Z^{-}_{(k+1)}, \vspace{-2pt}\\
         \textcolor{white}{\{~\boldsymbol{x}_{(k+1)},~c_{(k+1)}\}}
         \quad \boldsymbol{x}_{(k)},~c_{(k)},~\boldsymbol{u}^{+}_{(k)},~\boldsymbol{u}^{-}_{(k)} )\\
       \end{multlined}
     \end{equation*}
     \vspace{-10pt}
   \end{fleqn}%

   \State
   $ \boldsymbol{u}^{+}_{(k+1)} = \boldsymbol{u}^{+}_{(k)} + {\rho^{+}}(\text{vec}(W^{+}_{(k+1)} - Z^{+}_{(k+1)})$
   \State
   $ \boldsymbol{u}^{-}_{(k+1)} = \boldsymbol{u}^{-}_{(k)} + {\rho^{+}}(\text{vec}(W^{-}_{(k+1)} - Z^{-}_{(k+1)}) $
   \State
   $k = k + 1$
   \EndWhile
   \State\Return $W^{+}_{(k+1)}$, $W^{-}_{(k+1)}$,
   $Z^{+}_{(k+1)}$, $Z^{-}_{(k+1)}$, $\boldsymbol{x}_{(k+1)}$ and $c_{(k+1)}$
   \EndFunction
   \end{small}
 \end{algorithmic}
\end{algorithm}

%
In Algorithm~\ref{alg:ADMMupdate}, to solve for $W$, the problem boils down to
a regularized least squares problem and can be solved by gradient
descent method. 
To solve for $Z$, the problem boils down to a combination of a regularized
logistic regression problem and a least squares problem and can be solved by
gradient descent method. 
To solve for $\boldsymbol{x}$ and $c$, the problem boils down to 
a regularized logistic regression problem and can be solved by
gradient descent method.
The algorithm empirically converges. 
%

\subsection{Applying \model}
\label{sec:method:recommending} 
%
\model recommends drugs for each existing prescription $\boldsymbol{a}$
following a two-step procedure:
\begin{itemize}[noitemsep, leftmargin=*]
\item Step 1: use the learned \SLIM component to
  recommend a list of potential to-avoid/safe drug candidates $\{\drug_{r_j}\}$; 
\item Step 2: for each recommended candidate $\drug_{r_j}$,
  combine it with $\boldsymbol{a}$
  (i.e., $\boldsymbol{a} \cup \{\drug_{r_j}\}$) and
  use the learned \LogR component to predict the ADR probability
  of the new prescription. 
\end{itemize}
%
%
Note that in \LogR prediction, each recommended drug from \SLIM is considered
individually together with the prescription. 
It is also possible in real practice that multiple drugs all together should be
avoided as they, together with the prescription, may induce ADRs.
However, this problem is significantly more
difficult as the synergy of high-order drug combinations is highly non-trivial to
estimate. This problem will be tackled in our future research.

\subsubsection{Recommendation from \SLIM}
\label{sec:method:recommending:topn}

Given a prescription $\boldsymbol{a}$, the recommendation scores of all drugs with
respect to $\boldsymbol{a}$ are calculated as
$\tilde{ \boldsymbol{a}}^+ =  \boldsymbol{a}W^+$ and 
$\tilde{ \boldsymbol{a}}^- =  \boldsymbol{a}W^-$,
%
%
that is, $\tilde{ \boldsymbol{a}}^+$, the scores of being potential to-avoid drugs, 
are calculated from $W^+$,
and
$\tilde{ \boldsymbol{a}}^-$, the scores of being potential safe drugs, are calculated from
$W^-$.
Based on $\tilde{ \boldsymbol{a}}^+$ and $\tilde{ \boldsymbol{a}}^-$, the top-$M$ scored drugs
that are not in prescription $\boldsymbol{a}$ will be selected as to-avoid and 
safe drug candidates for $\boldsymbol{a}$, respectively.
%

\subsubsection{ADR Prediction and re-Ordering from \LogR}
\label{sec:method:recommending:pred}

The to-avoid and safe drug candidates from \SLIM are further
predicted using \LogR.
For a prescription $\boldsymbol{a}_i$, each of its candidates will be combined with $\boldsymbol{a}_i$
to form a new candidate to-avoid or safe prescription, denoted as $\boldsymbol{r}_i(j)^+$
(i.e., $\boldsymbol{r}_i(j)^+ = \boldsymbol{a}_i^+ \cup \{\drug_{r_j}^+\}$)
and $\boldsymbol{r}_i(j)^-$ (i.e., $\boldsymbol{r}_i(j)^- = \boldsymbol{a}_i^- \cup \{\drug_{r_j}^-\}$)
($j = 1, \cdots, N$), respectively.
Note that here $\boldsymbol{r}_i(j)^+$ and $\boldsymbol{r}_i(j)^-$ are both binary vectors. 
They have their corresponding score vectors from $\tilde{\boldsymbol{a}}^+$ and
$\tilde{\boldsymbol{a}}^-$, denoted as
$\boldsymbol{s}_i(j)^+$ and $\boldsymbol{s}_i(j)^-$, respectively.
Using ${\boldsymbol{f}_i(j)^+}$ to represent either $\boldsymbol{r}_i(j)^+$ or $\boldsymbol{s}_i(j)^+$,
and ${\boldsymbol{f}_i(j)^-}$ to represent either $\boldsymbol{r}_i(j)^-$ or $\boldsymbol{s}_i(j)^-$,
the ADR probabilities are calculated as $p( y^+ | {\boldsymbol{f}_i(j)^+}; \boldsymbol{x}, c)$
and $p( y^- | {\boldsymbol{f}_i(j)^-}; \boldsymbol{x}, c)$ (Equation~\ref{eqn:Probability}). 
%
%
That is, the to-avoid/safe drug candidates are only predicted for their labels as being true
to-avoid/safe drugs.
The drug candidates are then sorted based on their \LogR prediction scores, 
respectively, and the top-$N$ ranked drugs will be recommended.
Thus, the \SLIM step can be considered as an initial top-$M$ recommendation and the \LogR
step can be considered as a secondary re-ordering of initial candidates
for final recommendations. 
%
If $\boldsymbol{f}_i(j) = \boldsymbol{r}_i(j)$, the prediction method
is referred to as content-based prediction and denoted as \predBin.
If $\boldsymbol{f}_i(j) = \boldsymbol{s}_i(j)$, the prediction
method is referred to as score-based prediction and denoted as \predScr.

\section{Comparison Methods}
\label{sec:baseline}

\subsection{Random Method ({\Rand})}
\label{sec:baseline:random}

In the first comparison method, we fully randomly recommend drugs for each 
prescription as to-avoid drugs and safe drugs, respectively. 
This method is referred to as random method and denoted as \Rand.
\Rand serves as a baseline in which no learning is involved.

\subsection{Logistic Regression Only (\LROnly)}
\label{sec:Comparison:LROnly}

In this comparison method, only a \LogR model is learned to predict ADR labels.
Prescriptions that induce ADRs are used as positive
training data, and prescriptions that do not induce ADRs are used as negative
training data. No \SLIM models are learned in this method. 
For a prescription $\boldsymbol{a}$, each of all possible drugs that are not in
$\boldsymbol{a}$ is first combined with $\boldsymbol{a}$ to form a new prescription. 
Then \LogR is applied on the new prescription and produces a score of inducing ADRs.
The drugs that lead to the highest/lowest prediction scores are recommended as
the to-avoid/safe drugs.
This method is referred to as logistic regression only and denoted as \LROnly.

\subsection{Sparse Linear Method Only (\SLIMOnly)}
\label{sec:comparison:SLIMOnly}

In this method, a $\SLIM^+$ model is learned on ADR inducing prescriptions, and
a $\SLIM^-$ model is learned on non-ADR inducing prescriptions.
For a prescription, the top drugs recommended from $\SLIM^+$ are considered as to-avoid
drugs, and the top drugs recommended from $\SLIM^-$ are considered as safe drugs.
This method is referred to as \SLIM only and denoted as \SLIMOnly. 

\subsection{Separate Models (\Sep)}
\label{sec:comparison:seperate}

In this method, a \SLIM and a \LogR model are built separately, and used together
for recommendation.
Specifically, a $\SLIM^+$ and a $\SLIM^-$ as in \SLIM only are learned first. Meanwhile,
a \LogR model is also learned independently.
The top recommendations from $\SLIM^+$/$\SLIM^-$ are then predicted using the \LogR
model, and the top-scored drugs are recommended as to-avoid/safe drugs.
That is, the
recommendation process of this method is identical to that of \model. The difference
is that \SLIM and \LogR are learned independently, not jointly as in \model.
This method is referred to as separate \SLIM-\LogR model and denoted as \Sep.

%
\section{Materials}
\label{sec:materials}

\subsection{Mining Prescriptions}
\label{sec:materials:mining}
%
We first extracted high-order drug combinations from FDA Adverse Event Reporting System (FAERS)
\footnote{https://www.fda.gov/drugs/informationondrugs/ucm135151.htm}.
We use myopathy as the ADR of particular interest, and extract
64,892 case (myopathy) events, in which patients report myopathy after taking multiple drugs,
and 1,475,840 control (non-myopathy) events, in which patients do not report myopathy after
taking drugs.
Each of these events involves a combination of more than one drug. Each drug combination
is considered as a prescription. 
\begin{table}[!h]
  \caption{Contingency Table}
  \label{table:contigen}
    \vspace{-10pt}
    \centering
  \begin{threeparttable}
    \begin{small}
      \begin{tabular}{
	@{\hspace{10pt}}c@{\hspace{10pt}}|
	@{\hspace{10pt}}c@{\hspace{10pt}}          
	@{\hspace{10pt}}c@{\hspace{10pt}}|
	@{\hspace{10pt}}c@{\hspace{10pt}}
	}
	\toprule
        $\OR = \frac{n_1}{m_1}/ \frac{n_2}{m_2}$	& \adr			& no \adr		& total	\\
	\midrule
        $\boldsymbol{a}$					& $n_{1}$ 		& $m_{1}$		& $n_{1}+m_{1}$\\
         $\setminus\boldsymbol{a}$ 				& $n_{2}$     		& $m_{2}$ 		& $n_{2}+m_{2}$\\
	\midrule
	total					& $n_{1} + n_{2}$	& $m_{1} + m_{2}$	& $n_{1}+n_{2}+m_{1}+m_{2}$\\
	\bottomrule
      \end{tabular}
      \begin{tablenotes}
        \setlength\labelsep{0pt}
	\begin{footnotesize}
	\item
	  Here, $n_{1}$/$m_{1}$
          is the number of events where \D is taken and \adr does/doesn't occur; 
          $n_{2}$/$m_{2}$
          is the number of events where \D is not taken and \adr does/doesn't occur. 
          \par
	\end{footnotesize}
      \end{tablenotes}
    \end{small}
  \end{threeparttable}
    \vspace{-5pt}
\end{table}
%


%
Among all the involved prescriptions, 10,250 unique prescriptions
appear in both case and control events.
For those 10,250 prescriptions, we use Odds Ratio (\OR) 
to quantify their ADR risks.
The \OR for a prescription $\boldsymbol{a}$ is defined based on the contingency
table~\ref{table:contigen},
and is the ratio of the following two values:
1). the odds that the ADR occurs when $\boldsymbol{a}$ is taken
(i.e., $\frac{n_1}{m_1}$ in Table~\ref{table:contigen}); and
2). the odds that the ADR occurs when $\boldsymbol{a}$ is not taken
(i.e., $\frac{n_2}{m_2}$ in Table~\ref{table:contigen}).
\mbox{\OR $<$ 1} indicates the decreased risk of ADR after a patient takes the prescription,
\mbox{\OR$=1$} indicates no risk change, and \mbox{\OR $>$ 1} indicates the increased risk.
In the 10,250 prescriptions, 8,986 prescriptions have $\OR > 1$ and 1,264 prescriptions have
$\OR < 1$.
These two sets of prescriptions are denoted as \Mzero and \Nzero, respectively.
In addition to these prescriptions, there are 27,387 unique prescriptions
that only appear in case events and 621,449 unique prescriptions that only appear in control events.
These two sets are denoted as \Mplus and \Nminus, respectively.
The set of prescriptions in case events is denoted as \Mpos
(i.e., $\Mpos = \Mplus \cup \Mzero$) and is used as the positive set. 
The set of prescriptions in control events is
denoted as \Nneg (i.e., $\Nneg = \Nminus \cup \Nzero$) and is used as the negative set.
All these four sets together define a prescription dataset from FAERS, denoted as \Dall.
Table~\ref{table:Statis_whole} presents the statistics of \Dall.
\begin{table}[!h]
  \caption{Data Statistics}
  \label{table:Statis_whole}
  \vspace{-10pt}
  \centering
  \begin{threeparttable}
    \begin{small}
      \begin{tabular}{
	@{\hspace{8pt}}l@{\hspace{8pt}} 
	@{\hspace{8pt}}c@{\hspace{8pt}}
	@{\hspace{8pt}}r@{\hspace{8pt}}          
	@{\hspace{8pt}}r@{\hspace{8pt}}
        @{\hspace{4pt}}r@{\hspace{4pt}}
	@{\hspace{8pt}}r@{\hspace{8pt}}
	@{\hspace{8pt}}r@{\hspace{8pt}}
	}
	\toprule
        \multirow{2}{*}{dataset} & \multirow{2}{*}{stats} & \multicolumn{2}{c}{\Nneg} && \multicolumn{2}{c}{\Mpos} \\
        \cline{3-4} \cline{6-7}
	              & & \Nminus	& \Nzero	&& \Mzero        & \Mplus \\
	\midrule
	&\#$\{\D\}$     & 621,449	& 1,264		&& 8,986		& 27,387		\\
 	&\#$\{\drug\}$  & 1,209         & 417           && 881           & 1,201                 \\
\Dall	&avgOrd	& 6.100		& 2.351     	&& 3.588         & 7.096		\\
	&avgFrq	& 1.761		& 225.317      	&& 13.730        & 1.402		\\
	&avg\OR		& -	       	& 0.546		&& 16.343        & -	 		\\
	\cline{2-6}
	\midrule
        &\#$\{\D\}$	& 2,200         & 1,264		&& 2,464 	 & 1,000                 \\
        &\#$\{\drug\}$	& 562           & 417           &&  692           &   679                 \\
\Dataset   &avgOrd    & 2.678         & 2.351         && 3.809          & 7.615     \\
        &avgFrq        & 42.082        & 225.317       && 20.565         & 5.520        \\
        &avg\OR         & -             & 0.546         &&  31.998        & -        \\

	\bottomrule
      \end{tabular}
      \begin{tablenotes}
        \setlength\labelsep{0pt}
	\begin{footnotesize}
	\item         
	  Here, ``\#$\{\D\}$'' and ``\#$\{\drug\}$'' represent the number of prescriptions
          and the number of involved drugs, respectively.
	  In each set of prescriptions, ``avgOrd''/``avgFrq''/``avgOR''
          is the average number of drugs/average frequency/average \OR of all prescriptions. 
          \par
	\end{footnotesize}
      \end{tablenotes}
    \end{small}
  \end{threeparttable}
  \vspace{-5pt}
\end{table}
%


Note that the reason for myopathy as the only ADR is that myopathy has been well
studied and well understood compared to other ADRs
in terms of the drugs and drug mechanisms that induce myopathy.
Such knowledge will well support retrospective evaluation and validation on
drug recommendation with respect to myopathy.
On the other hand, multiple ADRs will introduce substantial difficulty to the recommendation
problem, and thus will be tackled in future research.

\subsection{Training Data Generation}
\label{sec:materials:training}

As shown in Table~\ref{table:Statis_whole}, \Mplus and \Mzero of \Dall
have fewer prescriptions than \Nminus and \Nzero, and the prescriptions in \Mplus
are very infrequent. 
To use more frequent and more confident prescriptions from case events, we further pruned
prescriptions from \Mplus and \Mzero of \Dall as follows.
From \Mplus of \Dall, we retained the top 1,000 most frequent prescriptions.
For \Mzero of \Dall, we applied right-tailed Fisher's exact test 
on the prescriptions
to test the significance of their \ORs at 5\% significance level
and retained prescriptions with statistically significant \ORs.
Thus, the pruned \Mplus and \Mzero contain statistically confident prescriptions,
which are very likely to induce myopathy, and therefore, these prescriptions are labeled as
positive. 

We retained all 1,264 the prescriptions in \Nzero of \Dall because this set is not large and contains
informative prescriptions that may or may not induce myopathy.
We further pruned \Nminus of \Dall and retained the top most frequent prescriptions.
The prescriptions from
\Nzero and the pruned \Nminus are labeled as negative. To make the positive and negative
training sets balance, we retained 2,200 prescriptions from \Nminus of \Dall.
The pruned dataset from \Dall is denoted as \Dataset.
Table~\ref{table:Statis_whole} presents the description of \Dataset.
Note that 
\Dataset is the set of labeled prescriptions that are used for model training. 
%
%

Note that the prescriptions in \Nneg do not induce
myopathy, but they may induce other ADRs because all prescriptions in 
FEARS are reported as inducing ADRs.
Therefore, \Nneg is not truly negative (i.e., free of ADRs) and only negative
with respective to myopathy.
Similarly, when safe drugs are referred to, they represent safety in terms of myopathy,
but not necessarily free of any ADRs.
%

\section{Evaluation Protocols and Metrics}
\label{sec:evaluation}

\subsection{Five-Fold Cross Validation}
\label{sec:evaluation:5fcv}

The performance of different methods is evaluated through five-fold cross validation.
The dataset \Dataset is randomly split into five folds of equal size 
(i.e., same number of prescriptions). 
Four folds are used for model training and the rest fold is used for testing.
In the fold for testing, one drug is randomly selected and removed from each prescription.
If the prescription after drug removal actually induces ADRs, it is removed from the
testing set. 
Then the to-avoid/safety drugs will be recommended for the rest of the prescription.
This is to make sure that it is the recommended to-avoid drugs that introduce DDIs and thus
ADRs.
This process is performed five times, with one fold for testing each time. The final result is
the average out of the five experiments.
In each split, the corresponding training set is denoted as \Dtrain, and the
testing set is denoted as \Dtest. 

\subsection{Knowledge Pool Generation}
\label{sec:evaluation:pool}

To evaluate the performance of various methods, we first define a knowledge pool of labeled
prescriptions, denoted as \DataPoll,
which the recommended new prescriptions can be searched from and evaluated against.
This knowledge pool serves as the entire knowledge space which is unknown during training. 
For a particular training set \Dtrain, \DataPoll can be either \mbox{$\Dataset \setminus \Dtrain$}
(i.e., \DataPoll is actually \Dtest in this case)
or \mbox{$\Dall \setminus \Dtrain$} (i.e., \DataPoll is much larger than \Dtest).
Note that \Dtrain is excluded from \DataPoll because
we can always search testing prescriptions in \Dtrain (i.e., known labeled prescriptions)
in order to identify corresponding to-avoid prescriptions
in \Dtrain, and therefore no recommendations will be needed.
By excluding \Dtrain from \DataPoll, we care about recommending particularly to-avoid/safe
drugs to prescriptions such the new prescriptions have not been observed before. We believe
this application is of more interest and significance in real practice. 


\subsection{Evaluation Metrics}
\label{sec:evaluation:metrics}

%
Each of the methods will generate $N$ recommended to-avoid/safe drugs for each
testing prescription. 
Given \DataPoll, the performance of these methods is evaluated using the following three
metrics.

\emph{Normalized Truncated Recall (\recall)} is denoted as \recall and defined as follows,
\begin{equation}
        \label{eqn:Recall}
        \rec = \frac{ \#\text{TP} }{\#\text{P} }, ~~
        \recall = \frac{\rec}{\max(\rec)},
\end{equation}
where $\#\text{TP}$ is the number of recommended to-avoid
prescriptions for testing prescriptions that appear in $\DataPoll^+$,
and $\#\text{P}$ is the number of all to-avoid prescriptions for testing
prescriptions in $\DataPoll^+$.
Note that the maximum possible value of \rec ($\max(\rec)$) can be smaller than 1. This is
because for each testing prescription, the number of its true to-avoid 
prescriptions existing in $\DataPoll^+$ (i.e., $\#\text{P}$) can be more than
the number of recommended prescriptions (i.e., $N$) that is allowed and therefore more than
$\#\text{TP}$.
Note that for each testing prescription, only top-$N$ to-avoid drugs are recommended, where
$N$ is a small number. 

\emph{Normalized Truncated Precision (\precision)} is denoted as \precision
and defined as follows
\begin{equation}
  \label{eqn:Precision}
  \precn = \frac{ \#\text{TP} }{\#\text{TP} + \#\text{FP} }, ~~
  \precision = \frac{\precn}{\max(\precn)},
\end{equation}
where $\#\text{FP}$ is the number of recommended to-avoid
prescriptions for testing prescriptions that are not in $\DataPoll^+$, and thus
$\#\text{TP} + \#\text{FP}$ is the number of total recommended to-avoid prescriptions. 
Note that the maximum possible value of \precn can be smaller than 1. This is
because for each testing prescription, the number of its true to-avoid 
prescriptions existing in $\DataPoll^+$ (i.e., $\#\text{P}$) can be smaller than
the number of recommended prescriptions (i.e., $N$) that is allowed,
and therefore $\#\text{TP}$ is smaller than $\#{\text{TP}} + \#\text{FP}$. 

\emph{Normalized Truncated Accuracy (\accuracy)} is denoted as \accuracy
and calculated as follows,
\begin{equation}
  \label{eqn:Accuracy}
  \acc = \frac{ \#\text{TP}+\#\text{TN} }{ \#\text{TP}+\#\text{TN} + \#\text{FP} + \#\text{FN}},~~
\accuracy = \frac{\acc}{\max(\acc)}, 
\end{equation}
where \#TN/\#FN is the number of recommended safe prescriptions that do
exist/do not exist in $\DataPoll^-$.
Note that the maximum possible value of \acc can be smaller than 1. This is
due to a similar reason as for $\precision < 1$, that is $\#\text{TP}$ can be smaller
than $\#\text{TP} + \#\text{FP}$, and $\#\text{TN}$ can be smaller than
$\#\text{TN} + \#\text{FN}$. 

Note that we did not use conventional ranking-based metrics such as average
precision at $k$ because we believe in clinical practice of multi-drug prescription,
people care whether all the to-avoid drugs can be correctly
recommended (i.e., a notion of recall)
much more than how the to-avoid drugs are ranked. 


%

%

\section{Experimental Results}
\label{sec:exp}

The presented experimental results in this section are more focused on to-avoid drug
recommendation due to space limit.
In addition, to-avoid drug recommendation is of more clinical significance compared
to safe drug recommendation.

\subsection{Method Comparison}
\label{sec:exp:overall}
%
\begin{table}[h!]
  \vspace{-5pt}
  \centering
  \caption{Comparison based on the Best \recall ($N=5$)}
  \label{table:recall:top5}
  \vspace{-10pt}    
  \begin{threeparttable}
      \begin{tabular}[]{
          @{\hspace{5pt}}l@{\hspace{5pt}}
	  @{\hspace{5pt}}l@{\hspace{5pt}}
          @{\hspace{5pt}}l@{\hspace{5pt}}
          @{\hspace{5pt}}r@{\hspace{5pt}}
          @{\hspace{5pt}}r@{\hspace{5pt}}
          @{\hspace{5pt}}r@{\hspace{5pt}}
        }
	\toprule
	\DataPoll	& mdl & prd & \recall   &  \precision  & \accuracy  \\
	\midrule								
\multirow{ 8}{*}{ \Dall } 	 	
        &        \multirow{2}{*}{\modelIn} &     \predScr	& \underline{\bf{{0.2361}}} & \bf{{0.2361}} & \bf{{0.2958}} \\
        &                                  &     \predBin	& \underline{\bf{{0.2361}}} & \bf{{0.2361}} & \bf{{0.2958}} \\
        &        \multirow{2}{*}{\modelEx} &     \predScr 	& 0.2108 & 0.2108 & 0.2795 \\
        &                                  &     \predBin 	& 0.2108 & 0.2108 & 0.2795 \\
        &                            \Rand &            - 	& 0.0060 & 0.0118 & 0.0238 \\
        &                          \LROnly &            - 	& 0.2082 & 0.2081 & 0.2430 \\
        &                        \SLIMOnly &            - 	& 0.2127 & 0.2126 & 0.2811 \\
        &                             \Sep &            - 	& 0.2127 & 0.2126 & 0.2813 \\
	\midrule
\multirow{ 8}{*}{ \Dataset }	
        &        \multirow{2}{*}{\modelIn} &     \predScr 	& \underline{\bf{0.2618}} & \bf{0.2615} & \bf{0.2815} \\
        &                                  &     \predBin 	& \underline{\bf{0.2618}} & \bf{0.2615} & \bf{0.2815} \\
        &        \multirow{2}{*}{\modelEx} &     \predScr 	& 0.2479 & 0.2479 & 0.2748 \\
        &                                  &     \predBin 	& 0.2479 & 0.2479 & 0.2748 \\
        &                            \Rand &            - 	& 0.0029 & 0.0059 & 0.0034 \\
        &                          \LROnly &            - 	& 0.1780 & 0.1781 & 0.1032 \\
        &                        \SLIMOnly &            - 	& 0.2469 & 0.2467 & 0.2754 \\
        &                             \Sep &            - 	& 0.2469 & 0.2467 & 0.2754 \\
	\bottomrule
      \end{tabular}
      \begin{tablenotes}
        \setlength\labelsep{0pt}
	\begin{footnotesize}
	\item
          The column ``mdl'' corresponds to models.
          The column ``prd'' corresponds
          to prediction methods.
          The best \recall is underlined.
          The best overall performance is \textbf{bold}. 
          \par
	\end{footnotesize}
      \end{tablenotes}
  \end{threeparttable}
  \vspace{-5pt}    
\end{table}
%

\begin{table}[h!]
  \vspace{-5pt}    
  \centering
  \caption{Comparison based on the Best \precision ($N=5$)}
  \label{table:precision:top5}
  \vspace{-10pt}
  \begin{threeparttable}
      \begin{tabular}[]{
          @{\hspace{5pt}}l@{\hspace{5pt}}
	  @{\hspace{5pt}}l@{\hspace{5pt}}
          @{\hspace{5pt}}l@{\hspace{5pt}}
          @{\hspace{5pt}}r@{\hspace{5pt}}
          @{\hspace{5pt}}r@{\hspace{5pt}}
          @{\hspace{5pt}}r@{\hspace{5pt}}
        }
	\toprule
	\DataPoll	& mdl & prd & \recall   &  \precision  & \accuracy  \\
	\midrule								
        \multirow{ 8}{*}{ \Dall }
        &        \multirow{2}{*}{\modelIn} &     \predScr	& {\bf{{0.2361}}} & \underline{\bf{{0.2361}}} & \bf{{0.2958}} \\
        &                                  &     \predBin	& {\bf{{0.2361}}} & \underline{\bf{{0.2361}}} & \bf{{0.2958}} \\

        &        \multirow{2}{*}{\modelEx} &     \predScr 	& 0.2108 & 0.2108 & 0.2795 \\
        &                                  &     \predBin 	& 0.2108 & 0.2108 & 0.2795 \\
        &                            \Rand &            - 	& 0.0060 & 0.0118 & 0.0238 \\
        &                          \LROnly &            - 	& 0.2082 & 0.2081 & 0.2430 \\
        &                        \SLIMOnly &            - 	& 0.2127 & 0.2128 & 0.2809 \\
        &                             \Sep &            - 	& 0.2127 & 0.2128 & 0.2811 \\
	\midrule
\multirow{ 8}{*}{ \Dataset }	
        &        \multirow{2}{*}{\modelIn} &     \predScr 	& \bf{0.2613} & \underline{\bf{0.2615}} & \bf{0.2821} \\
        &                                  &     \predBin 	& \bf{0.2613} & \underline{\bf{0.2615}} & \bf{0.2821} \\
        &        \multirow{2}{*}{\modelEx} &     \predScr 	& 0.2477 & 0.2479 & 0.2759 \\
        &                                  &     \predBin 	& 0.2477 & 0.2479 & 0.2759 \\
        &                            \Rand &            - 	& 0.0029 & 0.0059 & 0.0034 \\
        &                          \LROnly &            - 	& 0.1780 & 0.1781 & 0.1032 \\
        &                        \SLIMOnly &            - 	& 0.2465 & 0.2467 & 0.2754 \\
        &                             \Sep &            - 	& 0.2465 & 0.2467 & 0.2754 \\
	\bottomrule
      \end{tabular}
      \begin{tablenotes}
        \setlength\labelsep{0pt}
	\begin{footnotesize}
	\item
          The column ``mdl'' corresponds to models.
          The column ``prd'' corresponds
          to prediction methods.
          \mbox{The best \precision is underlined.
          The best overall performance is \textbf{bold}. }
          \par
	\end{footnotesize}
      \end{tablenotes}
  \end{threeparttable}
  \vspace{-5pt}
\end{table}
%

\begin{table}[t]
  \centering
  \caption{Comparison based on the Best \accuracy ($N=5$)}
  \label{table:accuracy:top5}
  \vspace{-10pt}
  \begin{threeparttable}
      \begin{tabular}[]{
          @{\hspace{5pt}}l@{\hspace{5pt}}
	  @{\hspace{5pt}}l@{\hspace{5pt}}
          @{\hspace{5pt}}l@{\hspace{5pt}}
          @{\hspace{5pt}}r@{\hspace{5pt}}
          @{\hspace{5pt}}r@{\hspace{5pt}}
          @{\hspace{5pt}}r@{\hspace{5pt}}
        }
	\toprule
	\DataPoll	& mdl & prd & \recall   &  \precision  & \accuracy  \\
	\midrule								
\multirow{ 8}{*}{ \Dall } 	 	
        &        \multirow{2}{*}{\modelIn} &     \predScr	& \bf{{0.2355}} & \bf{{0.2355}} & \underline{\bf{{0.2969}}} \\
        &                                  &     \predBin	& \bf{{0.2355}} & \bf{{0.2355}} & \underline{\bf{{0.2969}}} \\
        &        \multirow{2}{*}{\modelEx} &     \predScr 	& 0.2099 & 0.2099 & 0.2797 \\
        &                                  &     \predBin 	& 0.2108 & 0.2108 & 0.2797 \\
        &                            \Rand &            - 	& 0.0060 & 0.0118 & 0.0238 \\
        &                          \LROnly &            - 	& 0.2036 & 0.2039 & 0.2468 \\
        &                        \SLIMOnly &            - 	& 0.2019 & 0.2020 & 0.2830 \\
        &                             \Sep &            - 	& 0.2019 & 0.2020 & 0.2825 \\
	\midrule
\multirow{ 8}{*}{ \Dataset }	
        &        \multirow{2}{*}{\modelIn} &     \predScr 	& \bf{0.2589} & \bf{0.2592} & \underline{\bf{0.2832}} \\
        &                                  &     \predBin 	& \bf{0.2589} & \bf{0.2592} & \underline{\bf{0.2832}} \\
        &        \multirow{2}{*}{\modelEx} &     \predScr 	& 0.2477 & 0.2479 & 0.2765 \\
        &                                  &     \predBin 	& 0.2477 & 0.2479 & 0.2765 \\
        &                            \Rand &            - 	& 0.0029 & 0.0059 & 0.0034 \\
        &                          \LROnly &            - 	& 0.1780 & 0.1781 & 0.1032 \\
        &                        \SLIMOnly &            - 	& 0.2434 & 0.2432 & 0.2776 \\
        &                             \Sep &            - 	& 0.2434 & 0.2432 & 0.2776 \\
	\bottomrule
      \end{tabular}
      \begin{tablenotes}
        \setlength\labelsep{0pt}
	\begin{footnotesize}
	\item
          The column ``mdl'' corresponds to models.
          The column ``prd'' corresponds
          to prediction methods.
          The best \accuracy is underlined.
          The best overall performance is \textbf{bold}. 
          \par
	\end{footnotesize}
      \end{tablenotes}
  \end{threeparttable}
  \vspace{-10pt}    
\end{table}
%

%
Table~\ref{table:recall:top5}, ~\ref{table:precision:top5} and ~\ref{table:accuracy:top5}
present the performance of different methods in terms of their best \recall, best \precision
and best \accuracy, respectively, and the values on the other two evaluation metrics when
each respective best performance is achieved. Optimal parameters are not presented in the
tables due to space limit. 
In these tables, \model first recommends 20 drugs from its \SLIM component, and then its
\LogR component produces top-5 (i.e., $N=5$) scored drugs as to-avoid drug recommendations. 
In terms of best \recall as in Table~\ref{table:recall:top5}, when \Dall is used
as \DataPoll, 
\modelIn achieves the best performance (\recall = 0.2361), which is 11.0\%
better than the second best method \SLIMOnly and \Sep.
When \Dataset is used as \DataPoll, \modelIn also achieves the best performance
(\recall = 0.2618), which is 5.6\% better than the second best method \modelEx, and
6.0\% better than \SLIMOnly and \Sep. 
This indicates the power of joint learning of \SLIM recommendation component and
\LogR prediction component in \model. 
Meanwhile, 
\model performs at least 39-fold better than \Rand on \Dall as \DataPoll, and at least
90-fold better on \Dataset as \DataPoll. This indicates the strong ability of
\model in discovering true to-avoid drugs.
On the other hand, \LROnly is a strong baseline because it is at least 34-fold better and
61-fold better than \Rand in the two \DataPoll cases.
\SLIMOnly is even stronger than \LROnly in discovering true to-avoid drugs, as it is
2.2\% and 38.7\% better than \Rand in the two \DataPoll cases. 
However, when \LROnly or \SLIM is used alone, they are not any better than being used together
in \Sep, whereas the latter is still behind \model.

Very similar trends exist in Table~\ref{table:precision:top5} and ~\ref{table:accuracy:top5}.
In terms of best \precision as in Table~\ref{table:precision:top5}, \model outperforms the
second best methods \SLIMOnly and \Sep at 10.9\% and 6.0\%
in the two \DataPoll cases, respectively.
In terms of best \accuracy as in Table~\ref{table:accuracy:top5}, \model outperforms the
second best methods \SLIMOnly and \Sep at 4.9\% and 2.0\% in the two \DataPoll cases,
respectively. 
More importantly, \model achieves the overall
best \recall, best \precision and best \accuracy almost at the same time. For example,
when \modelIn achieves best \recall 0.2361 as in Table~\ref{table:recall:top5}, its
corresponding \precision 0.2361 is its best \precision as in Table~\ref{table:precision:top5},
and its corresponding \accuracy 0.2958 is same as its best \accuracy 0.2958 in
Table~\ref{table:accuracy:top5}. 
This indicates the strong capability of \model in accurately recovering true in-avoid drugs
and also true safe drugs. 

It is notable in Table~\ref{table:recall:top5}, ~\ref{table:precision:top5} and
~\ref{table:accuracy:top5}
that \modelIn consistently outperforms \modelEx in \recall, \precision and \accuracy.
\modelIn uses all the drugs and their scores recommended
by its \SLIM component in its \LogR component, whereas \modelEx only uses the drugs that
are already in corresponding prescriptions and their scores by \SLIM in
\LogR.
The use of all drugs from \SLIM in \LogR may enable \model to better learn its \SLIM
component and better capture the co-prescription relations among drugs that further lead
to their ADR labels. Meanwhile, it may generalize the capability of its \LogR component
to better predict new prescriptions, not to overfit on known prescriptions. 

It is also notable that the recommendation method \predBin and \predScr
(Section~\ref{sec:method:recommending:pred}) do not make a significant difference. They
almost achieve identical performance given a certain \model method as in
Table~\ref{table:recall:top5}, ~\ref{table:precision:top5} and ~\ref{table:accuracy:top5}.
This may be due to the fact that the \SLIM and \LogR components together
in \model are able to recover strong co-prescription patterns that closely
correlate to prescription ADR labels, and thus the patterns have their recommendation scores
close to binary. 
%

\subsection{Parameter Study}
\label{sec:exp:param}
%
\begin{table*} 
  \begin{minipage}{0.33\linewidth}
      \centering
      \caption{\recall of \modelIn+\predScr ($N=5$)}
        \vspace{-10pt}
  \label{table:param:recall}
  \begin{threeparttable}
      \begin{tabular}[]{
          @{\hspace{2pt}}l@{\hspace{2pt}}|
	  @{\hspace{2pt}}r@{\hspace{2pt}}
          @{\hspace{2pt}}r@{\hspace{2pt}}
          @{\hspace{2pt}}r@{\hspace{2pt}}
	  @{\hspace{2pt}}r@{\hspace{2pt}}
          @{\hspace{2pt}}r@{\hspace{2pt}}
        }
	\toprule
        {$\omega\backslash\alpha$}& {100} &  {50} &  {20} &  {10} &  {5} \\ 
	\midrule								
	20              & 0.2324 & 0.2344 & 0.2346 & 0.2349 & 0.2338 \\
	10              & 0.2315 & 0.2352 & 0.2349 & 0.2358 & 0.2344 \\
	5               & 0.2318 & 0.2355 & \textbf{0.2361} & \textbf{0.2361} & 0.2341 \\
	1		& 0.2315 & 0.2327 & 0.2321 & 0.2304 & 0.2290 \\
	\bottomrule
      \end{tabular}
      \begin{tablenotes}
        \setlength\labelsep{0pt}
	\begin{footnotesize}
	\item
          \DataPoll is \Dall. 
          The best performance is \textbf{bold}. 
          \par
	\end{footnotesize}
      \end{tablenotes}
  \end{threeparttable}
 \end{minipage}%
 \begin{minipage}{0.36\linewidth} 
  \centering
  \caption{\precision of \modelIn+\predScr ($N=5$)}
  \vspace{-10pt}
  \label{table:param:precision}
  \begin{threeparttable}
      \begin{tabular}[]{
          @{\hspace{2pt}}l@{\hspace{2pt}}|
	  @{\hspace{2pt}}r@{\hspace{2pt}}
          @{\hspace{2pt}}r@{\hspace{2pt}}
          @{\hspace{2pt}}r@{\hspace{2pt}}
	  @{\hspace{2pt}}r@{\hspace{2pt}}
          @{\hspace{2pt}}r@{\hspace{2pt}}
        }
	\toprule
        {$\omega\backslash\alpha$}& {100} &  {50} &  {20} &  {10} &  {5} \\ 
	\midrule								
	20              & 0.2325 & 0.2343 & 0.2347 & 0.2349 & 0.2339 \\
	10              & 0.2315 & 0.2351 & 0.2351 & 0.2359 & 0.2345 \\
	5               & 0.2321 & 0.2355 & \bf{0.2361} & 0.2359 & 0.2341 \\
	1		& 0.2315 & 0.2327 & 0.2321 & 0.2305 & 0.2290 \\
	\bottomrule
      \end{tabular}
      \begin{tablenotes}
        \setlength\labelsep{0pt}
	\begin{footnotesize}
	\item
          \DataPoll is \Dall. 
          The best performance is \textbf{bold}. 
          \par
	\end{footnotesize}
      \end{tablenotes}
  \end{threeparttable}
 \end{minipage}%
 \begin{minipage}{0.33\linewidth}
     \centering
  \caption{\accuracy of \modelIn+\predScr ($N=5$)}
  \vspace{-10pt}
  \label{table:param:accuracy}
  \begin{threeparttable}
      \begin{tabular}[]{
          @{\hspace{2pt}}l@{\hspace{2pt}}|
	  @{\hspace{2pt}}r@{\hspace{2pt}}
          @{\hspace{2pt}}r@{\hspace{2pt}}
          @{\hspace{2pt}}r@{\hspace{2pt}}
	  @{\hspace{2pt}}r@{\hspace{2pt}}
          @{\hspace{2pt}}r@{\hspace{2pt}}
        }
	\toprule
	{$\omega\backslash\alpha$}& {100} &  {50} &  {20} &  {10} &  {5} \\ 
	\midrule								
	20     & 0.2951 & 0.2919 & 0.2890 & 0.2895 & 0.2874 \\
	10     & 0.2943 & 0.2967 & 0.2938 & 0.2915 & 0.2891 \\
	5      & 0.2955 & \bf{0.2969} & 0.2958 & 0.2951 & 0.2924 \\
	1      & 0.2938 & 0.2953 & 0.2936 & 0.2919 & 0.2893 \\
	\bottomrule
      \end{tabular}
      \begin{tablenotes}
        \setlength\labelsep{0pt}
	\begin{footnotesize}
	\item
          \DataPoll is \Dall. 
          The best performance is \textbf{bold}. 
          \par
	\end{footnotesize}
      \end{tablenotes}
  \end{threeparttable}
 \end{minipage}%
  \vspace{-5pt}    
\end{table*}

Table~\ref{table:param:recall}, ~\ref{table:param:precision} and ~\ref{table:param:accuracy}
present the parameter study on $\omega$ and $\alpha$ as in Equation~\ref{opt:SLIMLRWsep}
on \modelIn with \predScr for top-5 (i.e., $N=5$) to-avoid drug recommendations. 
The pool \DataPoll is set to \Dall.
In \model, $\omega$ controls the weight of \LogR component
and $\alpha$ controls the Frobenius-norm regularizations in \SLIM component.
We don't present the parameter study over other parameters here. This is because
in our study, it is observed that \model requires a very small $\lambda$ (i.e.,
the parameter for the $\ell_1$-norm regularizations in \SLIM component), for example,
the optimal $\lambda$ is $10^{-6}$.
Also \model requires small 
$\beta$ and $\gamma$ (i.e., the parameters for \LogR component regularization), for
example, the optimal $\beta$ is $10^{-6}$ and the optimal $\gamma$ is $10^{-2}$ .
This indicates the very minor effects of these parameters on \model. 
We fix the optimal values for these parameters in Table~\ref{table:param:recall},
~\ref{table:param:precision} and~\ref{table:param:accuracy}, and study the
effects only from $\omega$ and $\alpha$.

Table~\ref{table:param:recall}, ~\ref{table:param:precision} and~\ref{table:param:accuracy}
demonstrate a very similar trend, that is, as $\omega$ becomes smaller (i.e., the weight
on \LogR component becomes lower in \model), all \recall, \precision and \accuracy
first increase and then decrease, and the optimal $\omega$ is around 5.
This represents a trade-off between the \SLIM component and the \LogR component in \model. 
Empirically more weight should be put on the \LogR component than on \SLIM component.
This may indicate that ADR label prediction is more difficult than co-prescription
pattern learning. 
Similarly, as $\alpha$ becomes smaller (i.e., the regularization on parameter $W$ of \SLIM
becomes less),  all \recall, \precision and \accuracy also 
first increase and then decrease, and the optimal $\alpha$ is around [10, 50].
Smaller $\alpha$ will introduce larger values in $W$ compared to larger $\alpha$, and thus
the relatively small optimal $\alpha$ indicates that $W$ captures strong patterns from
prescriptions.

\subsection{Top-$N$ Performance}
\label{sec:exp:topn}
%
\begin{table}[h!]
  \vspace{-5pt}    
  \centering
  \caption{Top-$N$ Performance of \modelIn with \predScr}
  \label{table:top}
    \vspace{-10pt}
  \begin{small}
  \begin{threeparttable}
      \begin{tabular}[]{
          @{\hspace{6pt}}l@{\hspace{6pt}}
          @{\hspace{6pt}}r@{\hspace{6pt}}
          @{\hspace{6pt}}r@{\hspace{6pt}}
          @{\hspace{6pt}}r@{\hspace{6pt}}
          @{\hspace{6pt}}r@{\hspace{6pt}}
          @{\hspace{6pt}}r@{\hspace{6pt}}
          @{\hspace{6pt}}r@{\hspace{6pt}}
          @{\hspace{6pt}}r@{\hspace{6pt}}
        }
	\toprule
	\multirow{2}{*}{$N$}
        & \multicolumn{ 3}{c}{ \Dall }    &
        & \multicolumn{ 3}{c}{ \Dataset } \\
        \cline{2-4} \cline{6-8}
	& \recall   &  \precision  & \accuracy &
        & \recall   &  \precision  & \accuracy  \\
	\midrule								
        \multirow{2}{*}{5} 
        & \bf{0.2358} & \bf{0.2359} & 0.2915      && \bf{0.2613} & \bf{0.2615} & 0.2821 \\
        & 0.2324      & 0.2325      & \bf{0.2929} && 0.2589      & 0.2592      & \bf{0.2832} \\
        \midrule
        \multirow{2}{*}{10} 
        & \bf{0.2497} & \bf{0.2497} & 0.3076      && \bf{0.3469} & \bf{0.3467} & 0.3419 \\
	& 0.2467      & 0.2467      & \bf{0.3110} && 0.3434      & 0.3432      & \bf{0.3439} \\
        \midrule
        \multirow{2}{*}{20} 
        & \bf{0.2885} & \bf{0.2887} & 0.2762      && \bf{0.4403} & \bf{0.4407} & 0.4016 \\
        & 0.2694      & 0.2695      & \bf{0.3289} && 0.4265      & 0.4273      & \bf{0.4135} \\
	\bottomrule
      \end{tabular}
      \begin{tablenotes}
        \setlength\labelsep{0pt}
	\begin{footnotesize}
	\item
          Column of ``N'' represents the number of recommended drugs. 
          Columns of ``\Dall'' and ``\Dataset'' represent that 
          ``\Dall'' and ``\Dataset'' are used as \DataPoll, respectively. 
          The \textbf{bold} performance is the best under the
          corresponding metrics in each column. 
          \par
	\end{footnotesize}
      \end{tablenotes}
  \end{threeparttable}
  \end{small}
  \vspace{-5pt}    
\end{table}
%

%
Table~\ref{table:top} presents the performance of the optimal \modelIn with \predScr
when $N=5, 10, 20$ to-avoid drugs are recommended.
As \modelIn recommends more to-avoid drugs, the performance in terms of all
\recall, \precision and \accuracy increases.
This indicates the capability of \modelIn in correctly ranking true to-avoid drugs in the
top of its recommendations. 

\subsection{Co-Prescription Patterns}
\label{sec:exp:patterns}

We consider $W^+$ and $W^-$ from \modelIn of the best performance
(i.e., $\omega = 5, \alpha = 20$).
$W^+$/$W^-$ captures the co-prescription patterns among drugs that often appear in
ADR inducing/no-ADR inducing prescriptions, respectively.
The learned $W^+$ has a density 3.6\%, and its top-5 largest values correspond
to the following drug pairs:
(desogestrel, ethinyl estradiol), (phenylephrine, hydrochlorothiazide),
(salmeterol, fluticasone propionate), (hydrocodone, acetaminophen), 
(ethinyl estradiol, etonogestrel).
These drug pairs are frequently prescribed together in \Mpos. 
The number of prescriptions in \Mpos that contain these drug pairs
is $38, 32, 44, 133$ and $38$, respectively, compared to the average number
of \Mpos prescriptions of all possible drug pairs, which is $1.22$.
Many of the co-prescribed drugs share similar medical purposes.
For example, salmeterol and fluticasone propionate are prescribed together 
to control and prevent symptoms caused by asthma or ongoing lung disease.
Ethinyl estradiol and etonogestrel are prescribed together for birth control, 
as well as ethinyl estradiol and desogestrel. 
Hydrocodone and acetaminophenare frequently are taken together to treat pains and reduce fever.

The learned $W^-$ has a density 0.8\%, and its top-5 largest values correspond
to the following drug pairs:
(lopinavir, ritonavir), (norelgestromin, ethinyl estradiol), (tenofovir, emtricitabine), 
(emtricitabine, tenofovir), (salmeterol, fluticasone propionate) 
%
These drug pairs are also frequently co-prescribed in \Nneg. 
The number of \Nneg prescriptions with these drugs pairs is
$36, 23, 35, 35$ and $35$, respectively, compared to $1.11$, the average number of
\Nneg prescriptions of all possible drug pairs. 
These drugs pairs also share similar medical purposes. 
For example, lopinavir and ritonavir, and tenofovir and emtricitabine are commonly used in HIV 
cocktail therapy.  
Norelgestromin and ethinyl estradiol are co-prescribed for birth control.
In the corresponding learned $\boldsymbol{x}$ (i.e., the parameter in the \LogR component),
the top-2 largest values correspond to drug simvastatin and atorvastatin.
These two drugs have been reported in SIDER Side Effect Resource~\footnote{http://sideeffects.embl.de} 
as to induce myopathy.
28.72\% 
of the ADR inducing prescriptions have at least one of these two drugs.

\subsection{Case Study}
\label{sec:exp:case}
%
\begin{table}[h!]
  \centering
  \caption{To-avoid Drug Recommendation Examples}
  \label{table:cases}
  \vspace{-10pt}
  \begin{small}
  \begin{threeparttable}
      \begin{tabular}{
	@{\hspace{3pt}}l@{\hspace{3pt}}
        @{\hspace{3pt}}p{0.7\linewidth}@{\hspace{0pt}}
        }
	\toprule
	recommendation  & prescription \\
	\midrule				
				
        \textbf{pramipexole}      & carbidopa, \textbf{citalopram}, alprazolam, levothyroxine, entacapone, clonazepam, \textbf{atorvastatin}, levodopa\\
        \textbf{rosuvastatin}     & \textbf{niacin}, \textbf{simvastatin}\\
        \midrule
        \textbf{phenytoin}        & oxcarbazepine, levetiracetam\\
        \textbf{simvastatin}      & ramipril, ranolazine, amlodipine, trimethoprim, nitroglycerin, acetylsalicylic acid, isosorbide mononitrate, carvedilol, nicorandil\\
        \midrule
        abacavir                  & \textbf{nelfinavir}, \textbf{tenofovir}, \textbf{zidovudine}, \textbf{ritonavir}, stavudine, \textbf{lamivudine}, \textbf{didanosine}, \textbf{zalcitabine}, lopinavir\\
        rosiglitazone             & metformin, \textbf{simvastatin}\\
        \midrule
        alendronate               & vitamin c, triamterene, valdecoxib, lisinopril, desloratadine, hydrochlorothiazide, anastrozole\\
        risperidone               & valproic acid, clozapine\\
	\bottomrule
      \end{tabular}
      \begin{tablenotes}
        \setlength\labelsep{0pt}
	\begin{footnotesize}
	\item
	Drugs that are reported in SIDER to induce myopathy are \textbf{bold}.
          \par
	\end{footnotesize}
      \end{tablenotes}
  \end{threeparttable}
  \end{small}
    \vspace{-10pt}    
\end{table}
%



        %

%
Table~\ref{table:cases} presents some examples of prescriptions and their
recommended to-avoid drugs from \model such that the corresponding new prescriptions
are true positive. Drugs that are reported in SIDER to induce myopathy are bold.
Based on whether the prescriptions have myopathy-inducing drugs or not, and whether
the recommended drugs induce myopathy or not, all the new prescriptions can be classified
into four categories (corresponding to the four row blocks in Table~\ref{table:cases}).
Note that in the last category, even the prescriptions do not contain any myopathy-inducing
drugs, \model is able to recommend other non-ADR inducing drugs such that the new
prescriptions will induce myopathy.
This demonstrates the strong capability of \model for co-prescription pattern learning and
ADR label prediction.
%

\section{Discussions and Conclusions}
\label{sec:disc}

In this paper, we formally formulated the to-avoid drug recommendation problem
and the safe drug recommendation problem.
We developed a \name (\model) to tackle the recommendation problems, and developed
real datasets and new evaluation protocols to evaluate the model.
Our experiments demonstrate strong performance of \model compared to other methods.

In \model, we use simplistic linear models. Non-linear models (for both recommendation
and label prediction) together may be able to better discover co-prescription patterns
and predict labels jointly. We will investigate non-linear models in the future work.
Also it is common that DDIs induce multiple ADRs simultaneously. This problem is substantially
more difficult but deserve future investigation.

\section*{Acknowledgment}
\label{sec:acknowledgment}

This material is based upon work supported by the National Science Foundation
  under Grant Number IIS-1566219 and and IIS-1622526.

\bibliographystyle{ACM-Reference-Format}
\bibliography{paper}


\end{document}